\documentclass[preprint,aps,amssymb,showpacs,superscriptaddress,nofootinbib]{revtex4}
\usepackage{graphicx}
\newcommand{\del}{\partial}

\begin{document}

\preprint{}

\title{ Quantum realizations of Hilbert-Palatini second-class
  constraints}

\author{Sandipan Sengupta} \email{sandi@imsc.res.in}
\affiliation{The Institute of Mathematical Sciences\\
CIT Campus, Chennai-600 113, INDIA.}

\begin{abstract}
  In a classical theory of gravity, the Barbero-Immirzi parameter
  ($\eta$) appears as a topological coupling constant through the
  Lagrangian density containing the Hilbert-Palatini term and the
  Nieh-Yan invariant. In a quantum framework, the topological
  interpretation of $\eta$ can be captured through a rescaling of the
  wavefunctional representing the Hilbert-Palatini theory, as in the
  case of the QCD vacuum angle.
  However, such a rescaling cannot be realized for pure gravity within
  the standard (Dirac) quantization procedure where the second-class
  constraints of Hilbert-Palatini theory are eliminated beforehand.
 
  Here we present a different treatment of the Hilbert-Palatini
  second-class constraints in order to set up a general rescaling
  procedure (a) for gravity with or without matter and (b) for any
  choice of gauge (e.g.  time gauge). The analysis is developed using
  the Gupta-Bleuler and the coherent state quantization methods.

\end{abstract}

\pacs{04.20.Fy, 04.20.Gz, 04.60.Ds}

\maketitle{}

\section{Introduction}
It has been suggested \cite{gambini} that the Barbero-Immirzi
parameter ($\eta$) as appearing in the Holst action \cite{holst}
should emerge as a topological parameter in the formally quantized
theory of gravity, much like the $\theta$ parameter in QCD
\cite{gambini,jackiw}. However, the Holst term is not a topological
density. This implies that one cannot think of an analogue of the
Chern-Simons functional as in QCD, through which the non-perturbative
vacuum structure of gravity, as anticipated, can manifest itself.
Hence within this formulation, $\eta$, the coefficient of the Holst
term does not qualify for a topological interpretation.

In a classical context, the topological nature of Barbero-Immirzi
parameter was clarified in ref.\cite{kaul} through the Hamiltonian
analysis of the following action :
\begin{equation} \label{ny}
  L~=~\frac{1}{2}e\Sigma^{\mu\nu}_{IJ}R_{\mu\nu}^{IJ}~+~\frac{\eta}{2}I_{NY}
\end{equation} 
with
$\Sigma_{IJ}^{\mu\nu} ~ := ~
\frac{1}{2}~(e_{I}^{\mu}e_{J}^{\nu}-e_{J}^{\mu}e_{I}^{\nu}) ~,~ R^{~~~
  IJ}_{\mu\nu}(\omega) ~ := ~ \del_{[\mu} \omega_{\nu]}^{~IJ} +
\omega_{[\mu}^{~IK}\omega_{\nu]K}^{~ ~ ~J}$.  Here, the first term is
the Hilbert-Palatini term. In the second term, $\eta$ is a coefficient
of the Nieh-Yan invariant $I_{NY}$, which is a topological density
\cite{nieh} , that is, a total divergence :
\begin{eqnarray*}
  I_{NY}&~=~&\del_{\mu}[\epsilon^{\mu\nu\alpha\beta}e^{I}_{\nu}D_{\alpha}e_{I\beta}]\\
  &~=~&\del_{\mu}[\epsilon^{\mu\nu\alpha\beta}e^{I}_{\nu}(\del_{\alpha}e_{I\beta}+\omega_{\alpha IJ}e_{\beta}^{J})]\\
  &~=~&\del_{\mu}j_{NY}^{\mu}(e,\omega)
\end{eqnarray*}
This indicates that the two different canonical formulations, i.e.,
the ones with and without the Barbero-Immirzi parameter, are related
through a canonical transformation generated by $j_{NY}^{t}$.

However, the fact that $I_{NY}$ is a topological density suggests that
there must exist `large gauge transformations' under which $\int d^3
xj_{NY}^{t}$ transforms non-trivially \cite{zanelli}. This would imply
the existence of different inequivalent vacua parametrised by $\eta$
in the quantum theory of gravity. Then the canonical transformation as
mentioned above cannot be implemented unitarily in the corresponding
quantum theory. In this context, it is worthwhile to note that the
appearance of $\eta$ in the spectrum of the area operator
\cite{ashtekar} also strongly hints at such a possibility.

In a quantum framework, this issue of a non-unitary realization of the
canonical transformation generated by $j^t_{NY}$ can be studied
through the procedure of rescaling. We recall that in the case of QCD,
the interpretation of $\theta$ as a vacuum angle can be understood
from this perspective alone, i.e., through the rescaling of the
wavefunction $\Psi$ representing the quantum theory \cite{jackiw}:
\begin{eqnarray}\label{qcdwavefn} \Psi'~=~e^{i\theta\int d^{3}x Y_{cs}}~\Psi
\end{eqnarray} where $Y_{cs}$ is the Chern-Simons functional. Under large gauge transformations $\Psi$ changes by a phase which is cancelled by the phase picked up by the exponential operator, leaving $\Psi'$ to be invariant. Thus the operator $e^{i\theta\int d^{3}x Y_{cs}}$ essentially implements the
large gauge transformations. Here $\theta$ is manifestly an angular parameter characterising the inequivalent
sectors in the quantum theory. Through this rescaling one can induce a canonical transformation on the $\theta$-independent canonical operators (characterising the standard QCD Lagrangian density without any $\theta$ term) to arrive at the $\theta$-dependent canonical operators (corresponding to the Lagrangian density containing the $\theta$ term). Note that the canonical formulation thus obtained corresponds precisely to  the one coming from a Hamiltonian analysis of the QCD Lagrangian density which contains the $\theta$ term (i.e., the Pontryagin density). 

Here we attempt to develop a similar rescaling procedure for gravity
{\em with or without matter}. Such a construction is particularly
relevant in order to study the vacuum structure of gravity. This would
also allow one to go from the canonical formulation without $\eta$
(Hilbert-Palatini theory) to the one containing $\eta$ (corresponding
to the Lagrangian density in (\ref{ny})).  To proceed, let us consider
the rescaling of the wavefunctional $\Psi$ representing the formally
quantized Hilbert-Palatini theory:
\begin{eqnarray}\label{wavefn} \Psi'~=~e^{i\eta\int d^{3}x
    \hat{Y}(q)}~\Psi(q)
\end{eqnarray} where q's are the coordinates in the quantum configuration
space and Y, the rescaling functional is given by :
\begin{eqnarray} \label{y} Y&~=~&\frac{1}{2}~j_{NY}^{t}(e,\omega)
\nonumber\\ &~=~&\frac{1}{2}~\epsilon^{abc}e^{I}_{a}D_{b}e_{Ic}
\end{eqnarray}
However, as demonstrated in the next section, if the Dirac procedure
is used to eliminate the second-class constraints before quantization
\cite{dirac}, the operator $\hat{j}_{NY}^{t}$ vanishes `strongly' (in
the sense of Dirac in \cite{dirac}). This reduces the exponential in
(\ref{wavefn}) to an identity operator which clearly cannot implement
the corresponding `large gauge transformations'. Thus in Dirac's
method of quantization, there is no way for $\eta$ to emerge as a
vacuum angle through the rescaling as in QCD. This also means that one
cannot arrive at the canonical constraints containing $\eta$ starting
from the Hilbert-Palatini theory via the rescaling route.

We note that the procedure of rescaling has been applied earlier to
gravity coupled to spin-$\frac{1}{2}$ fermions \cite{mercuri}.
However, the approach in ref.\cite{mercuri} uses Dirac's method to
solve the second-class constraints {\em before} quantization, or in
other words, uses the connection equation of motion. As explained
already, this method cannot be applied to {\em pure} gravity, since
the rescaling functional Y in (\ref{y}) vanishes when the connection
equation is used.

Thus one is forced to adopt a different quantization procedure other
than the standard one of Dirac in order to set up a general rescaling
method for gravity {\em with or without matter}. In this paper we
address this issue. Here we reformulate the Hilbert-Palatini canonical
constraints using the Gupta-Bleuler and the coherent state
quantization methods which do not require the elimination of
second-class constraints to begin with. This is then followed by a
demonstration of the rescaling which essentially leads to the real
Ashtekar-Barbero canonical formulation.

The general idea of the Gupta-Bleuler quantization \cite{glikman} is
to split the original set of first and second-class constraints into a
holomorphic and an anti-holomorphic set of first-class constraints
related through the hermitian conjugation. The physical subspace
contains only those ket states which are annihilated by the
holomorphic set. Here we apply this method to Hilbert-Palatini theory.
The resulting space of physical wavefunctions is then used to employ
the rescaling. This leads to the canonical formulation based on
(\ref{ny}), as desired.  Next we repeat the exercise using the
coherent state quantization for constrained systems\cite{klauder}.
There we consider a squared sum of the original second-class
constraints to define the physical Hilbert space. Note that such
squared combinations also appear in the context of the Master
constraint programme\cite{thiemann}, where the constraints are
enforced in a different way than above.

In contrast to Dirac's approach, our analysis in either cases does not
require the use of the connection equation of motion for the
rescaling.  This particular feature is essential in order to recover a
complete topological interpretation of the Barbero-Immirzi parameter,
independent of any matter coupling.

In the following section we demonstrate the rescaling procedure in
time gauge, first in the Gupta-Bleuler and then in the Coherent state
approach. We work in a representation diagonal in the densitized triad
operators, as is required. In section III, we generalise our
construction without fixing any gauge.  Section IV contains a few
concluding remarks.

\section{Rescaling in time gauge}

The Hilbert-Palatini Lagrangian density is given by :
\begin{equation} \label{HP}
  L~=~\frac{1}{2}e\Sigma^{\mu\nu}_{IJ}R_{\mu\nu}^{IJ}
\end{equation} 
We parametrize the tetrad fields as \cite{kaul} :
\begin{equation}
e^{I}_{t} ~ = ~ \sqrt{eN}M^{I}+N^{a}V_{a}^{I} ~ , ~ e^{I}_{a} ~ = ~
V^{I}_{a} ~ ~ ~;~ ~ ~ M_{I}V_{a}^{I} ~ = ~ 0 ~ , ~ M_{I}M^{I} ~ = ~ -1
\end{equation} 
and then the inverse tetrad fields are : 
\begin{eqnarray}
e^{t}_{I} ~ = ~ -\frac{M_{I}}{\sqrt{eN}} ~ , ~ e^{a}_{I} ~ = ~
V^{a}_{I}+\frac{N^{a}M_{I}}{\sqrt{eN}}  ~ ~; \nonumber \\
M^{I}V_{I}^{a} ~ = ~ 0 ~,~ V_a^I V^b_I ~ = ~ \delta_a^b ~,~ V_a^I
V^a_J ~ = ~ \delta^I_J + M^IM_J 
\end{eqnarray} 
Introducing the fields 
\begin{eqnarray} \label{tomega}
  E^{a}_{i}~=2e\Sigma^{ta}_{0i}~~,~~\chi_{i}~=~\frac{M_{i}}{M_{0}}~~,~~
  \tilde{\omega}^{~0i}_{b}~=~\omega_{b}^{~0i}~-~\chi_{m}\omega_{b}^{~im}
  ~~,~~\zeta_j~=~\omega_{aij}\chi^i
\end{eqnarray} 
the Lagrangian density in (\ref{HP}) can be written as :
\begin{equation} \label{hp}
  L~=~E^{a}_{i}\del_{t}\tilde{\omega}_{a}^{~0i}~+~\zeta^{i}\del_{t}\chi_{i}~-~\omega_{t}^{0i}G_{0i}~-~\frac{1}{2}~\omega_{t}^{ij}G_{ij}~-~NH~-~N^{a}H_{a}
\end{equation}
$H ,~H_a$ and $G^{boost}_{i}$ and $G^{rot}_{i}$ are the scalar, vector,
boost and rotation constraints, respectively ($G_i^{\mathrm{boost}} :=
G_{0i}\, ,\ G_i^{\mathrm{rot}} := \frac{1}{2} \epsilon_{ijk} G^{jk}$)
.

In terms of the canonical variables, the Nieh-Yan functional in
(\ref{y}) becomes :
\begin{eqnarray} \label{ybig}
  Y~&=&~\frac{1}{4}~\sqrt{E}\epsilon^{ijk}E^{a}_{j}E^{b}_{k}~~\left[-~\chi_{i}\del_{a}\left(\sqrt{E}\epsilon^{lmn}\epsilon_{bcd}~\chi_{l}E^{c}_{m}E^{d}_{n}\right)
    ~+~\del_{a}\left(\sqrt{E}\epsilon^{imn}\epsilon_{bcd}E^{c}_{m}E^{d}_{n}\right)~\right]\nonumber \\
  &~&-~\frac{1}{2}~M_{kk}-~\frac{1}{2(1-\chi^2)}\chi_{k}\chi_{l}M_{kl}
  ~-~\epsilon^{ijk}\chi_{j}\tilde{\omega}^{~0i}_{b}E^{b}_{k}
\end{eqnarray}
where we have used the identities
\begin{eqnarray*}
\epsilon^{abc}V_{kc}~=~\sqrt{E}\epsilon_{ijk}E^{ai}E^{bj},~~  
\sqrt{E}V_{ck}~=~E_{ck}~~\mathrm{and~~} \sqrt{E}V_{c0}~=~\chi^{k}E_{ck}
\end{eqnarray*}
and also the decomposition\footnote{The parametrisation for
  $\omega_{aij}$ here is different from that in ref.\cite{kaul}}:
\begin{eqnarray*}
  \omega_{aij}~=~\frac{1}{2}E_{a[i}\zeta_{j]}~+~\frac{1}{2(1-\chi^2)}\epsilon_{ijk}E_{al}M^{kl},~~
  M^{kl}~=~M^{lk}
\end{eqnarray*}
which is just a way to represent the nine components of $\omega_{aij}$
in the basis of three $\zeta_{i}$'s and six $M_{kl}$'s.

One can choose the time gauge by putting $\chi_{i}\approx0$ . As this
condition forms a second class pair with the boost constraint, they
have to be imposed together.

The boost constraint is given by -
\begin{eqnarray*}
  G^{boost}_{i}&~=&~-~\del_{a}E^{a}_{i}~-~\omega_{aij}E^{a}_{j}\\
  &~=&~-~\del_{a}E^{a}_{i}~+~\zeta_{i}
\end{eqnarray*}
which is solved by the condition 
\begin{eqnarray} \label{zeta}
\zeta_{i}=\del_{a}E^{a}_{i}~~~~.
\end{eqnarray}

The first-class set of constraints are given by the following
expressions :
\begin{eqnarray} \label{rot}
  G^{rot}_{i}&~=~&\epsilon_{ijk}\omega_{a}^{0j}E^{a}_{k} \nonumber\\
  H_{a}&~=~&E^{b}_{i}R^{~0i}_{ab} \nonumber\\
  &~=~&E^{b}_{i}\del_{[a}\omega_{b]}^{0i}~-~\omega_{a}^{0i}\zeta_{i}~+~[~\epsilon_{ijl}E_{aj}\zeta_{i}~-~E_{ak}M^{kl}~]~G^{rot}_l \nonumber\\
  H&~=~&-~\frac{1}{2}~E^{a}_{i}E^{b}_{j}R^{~ij}_{ab}   \nonumber\\
  &~=~&E^{a}_{i}\del_{a}\zeta_{i}~+~\frac{1}{2}~\zeta_{i}\zeta_{i}~-~\frac{1}{2}~E^{a}_{i}E_{bj}\zeta_{i}\del_{a}E^{bj}~+~\frac{1}{2}\epsilon_{ijm}~E^{a}_{i}E_{bn}\del_{a}E^{bj}M^{mn} \nonumber\\
  &~&~+~\frac{1}{8}~[~2\zeta_{i}\zeta_{i}~+~M^{kk}M^{ll}~-~M^{kl}M^{kl}]~-~\frac{1}{2}E^{a}_{i}E^{b}_{j}~\omega_{[a}^{0i}\omega_{b]}^{0j}
\end{eqnarray}
where $\zeta_{i}$ is given by (\ref{zeta}) .

From (\ref{ybig}) , we get the following expression for Y in time
gauge :
\begin{eqnarray} \label{ytime}
  Y&~=~&\frac{1}{4}\sqrt{E}\epsilon^{ijk}E^{a}_{j}E^{b}_{k}\del_{a}\left(\sqrt{E}\epsilon^{imn}\epsilon_{bcd}E^{c}_{m}E^{d}_{n}\right)~-~\frac{1}{2}M_{kk}
\end{eqnarray}

\subsection{Classical second-class constraints} 

As the Lagrangian density (\ref{hp}) is independent of the velocities
associated with $M_{kl}$, we have the primary constraints involving
the corresponding momenta :
\begin{eqnarray} \label{pikl} 
\pi_{kl}~\approx~0~
\end{eqnarray}
These in turn imply secondary constraints, which essentially lead to
the vanishing of torsion (see \cite{kaul} for details) :
\begin{eqnarray} \label{mkl}
  &&~ [H,\pi_{kl}]~\approx~0~ \nonumber \\
  ~=>&&~ \epsilon_{ijk}E^{a}_{i}E_{bl}\del_{a}E^{b}_{j}~+~\frac{1}{2}~(~M^{ii}\delta_{kl}~-~M_{kl}~)~+~(k\leftrightarrow l)\approx~0\nonumber\\
  ~=>&&~M_{kl}~-~F_{kl}(E^{a}_{i})~\approx~0
\end{eqnarray}
where we have defined $F_{kl}$ as \footnote{In the presence of matter
  coupling this would be
  $~[~M_{kl}-F_{kl}(E^{a}_{i},\phi_{m})~]\approx~0~$, where $\phi_{m}$
  denotes the matter fields, eg. fermions when gravity is coupled to
  fermions} :
\begin{eqnarray} \label{fkl} 
  F_{kl}(E^{a}_{i})~=~\frac{1}{2}~\epsilon_{ijm}E^{a}_{i}E^{b}_{m}\del_{a}E_{bj}~\delta_{kl}~-~\epsilon_{ijk}E^{a}_{i}E^{b}_{l}\del_{a}E_{bj}~+~(k\leftrightarrow l)
\end{eqnarray}

Dirac's prescription leads to the next step where the second-class
constraints are solved before quantization or are eliminated through
Dirac brackets. This is equivalent to imposing them `strongly' as
operator conditions\cite{dirac}. Thus the physical subspace of the
original Hilbert space would be obtained through the states which are
annihilated by the operators corresponding to the remaining set of
first-class constraints. However, here the second-class pair in
(\ref{pikl}) and (\ref{mkl}), when enforced strongly, leads to the
vanishing of the rescaling functional Y. Thus, the Dirac quantization
procedure as it is cannot provide any passage to the new set of
constraints corresponding to the Lagrangian density containing the
Nieh-Yan term.

Hence, one must adopt alternative quantization procedures to impose
the second-class constraints in the quantum state space. Here we first
employ a method which is a slight generalisation of the Gupta-Bleuler
approach in electrodynamics, and then repeat the exercise using the
coherent state quantisation. Both the cases result in a non-vanishing
rescaling functional through which the canonical transformation can be
carried out.

\subsection{Gupta-Bleuler quantization} 

Following the general idea of Gupta-Bleuler quantization, we have to
find suitable holomorphic and anti-holomorphic sets containing all the
constraints. Here the sets can be defined as:
\begin{eqnarray}\label{set}
  C&:=&\left(G^{i}_{rot},~ H_a, ~\tilde{H},~(Q_{kl}+i\alpha\pi_{kl})\right)\nonumber\\
  C^{\dagger}&:=&\left(G^{i}_{rot},~ H_a, ~\tilde{H}^{\dagger},~(Q_{kl}-i\alpha\pi_{kl})\right)
\end{eqnarray}
where $\alpha$ is a constant and
$\tilde{H},~\tilde{H}^{\dagger},~Q_{kl}$ are defined as:
\begin{eqnarray*}
  \tilde{H}&~=~&E^{a}_{i}\del_{a}\zeta_{i}~+~\frac{1}{2}~\zeta_{i}\zeta_{i}~-~\frac{1}{2}~E^{a}_{i}E_{bj}\zeta_{i}\del_{a}E^{bj}~+~\frac{1}{2}\epsilon_{ijm}~E^{a}_{i}E_{bn}\del_{a}E^{bj}(M_{mn}+i\alpha\pi_{mn}) \nonumber\\
  &+&\frac{1}{8}~[~2\zeta_{i}\zeta_{i}~+~(M_{kk}+i\alpha\pi_{kk})(M_{ll}+i\alpha\pi_{ll})~-~(M_{kl}+i\alpha\pi_{kl})(M_{kl}+i\alpha\pi_{kl})]\nonumber\\
  &-&\frac{1}{2}E^{a}_{i}E^{b}_{j}~\omega_{[a}^{0i}\omega_{b]}^{0j}\\
  \tilde{H}^{\dagger}&~=~&E^{a}_{i}\del_{a}\zeta_{i}~+~\frac{1}{2}~\zeta_{i}\zeta_{i}~-~\frac{1}{2}~E^{a}_{i}E_{bj}\zeta_{i}\del_{a}E^{bj}~+~\frac{1}{2}\epsilon_{ijm}~E^{a}_{i}E_{bn}\del_{a}E^{bj}(M_{mn}-i\alpha\pi_{mn}) \nonumber\\
  &+&\frac{1}{8}~[~2\zeta_{i}\zeta_{i}~+~(M_{kk}-i\alpha\pi_{kk})(M_{ll}-i\alpha\pi_{ll})~-~(M_{kl}-i\alpha\pi_{kl})(M_{kl}-i\alpha\pi_{kl})]\nonumber\\
  &-&\frac{1}{2}E^{a}_{i}E^{b}_{j}~\omega_{[a}^{0i}\omega_{b]}^{0j}\\
  Q_{kl}&~=~&M_{kl}-F_{kl}(E^{a}_{i})\nonumber
\end{eqnarray*} 
Thus we have two sets of first-class constraints which satisfy the
required algebra given by\cite{glikman}:
\begin{eqnarray}\label{algebra} 
  &~&[C_A,C_B]~\approx~0 ~\approx~[C^{\dagger}_A,C^{\dagger}_B], \nonumber\\
  &~&[C_A,C^{\dagger}_B]~\approx~ Z_{AB}
\end{eqnarray}
where $Z_{AB}$, the central charge is a function of $\alpha$ in our
case.

In the definitions (\ref{set}), instead of $H$, one needs to take the
classically equivalent constraints $\tilde {H}$ and $\tilde
{H}^{\dagger}$ in order to ensure the abelian property of the
individual sets, and hence to reproduce the correct algebra as in
(\ref{algebra}). $\tilde {H}$ and $\tilde {H}^{\dagger}$ are obtained
by replacing $M_{kl}$ in $H$ by $(M_{kl}+i\alpha\pi_{kl})$ and
$(M_{kl}-i\alpha\pi_{kl}$) , respectively.

Next we define a representation based on the fundamental commutation
relations as:
\begin{eqnarray*} 
  && \hat{E}^{a}_{i}~|\Psi>~=~E^{a}_{i}~|\Psi>~~,~~\hat{\omega}^{~0i}_{a}~|\Psi>~=~-~i\frac{\delta}{\delta \hat{E}^{a}_{i}}~|\Psi> \\
  && \hat{M}_{kl}~|\Psi>~=~M_{kl}~|\Psi>~~,~~\hat{\pi}^{kl}~|\Psi>~=~-~i\frac{\delta}{\delta \hat{M}_{kl}}~|\Psi>
\end{eqnarray*}
Here $|\Psi>$ represents the formally quantized Hilbert-Palatini
theory.  The physical subspace is obtained through the realisation of
the set $C$ on the ket states:
\begin{eqnarray}
\hat{C}~|\Psi>~=~0
\end{eqnarray}
Thus, the constraints involving the canonical pair
($\hat{M}_{kl},~\hat{\pi}_{kl})$ act as:
\begin{eqnarray}\label{annihilation} 
  (\hat{Q}_{kl}+i\alpha\hat{\pi}_{kl})~|\Psi>~=~0
\end{eqnarray}
The hermitian conjugation of the above implies that the physical bra
states are annihilated by $C^{\dagger}$.

From (\ref{annihilation}), it follows that the original second-class
constraints are satisfied individually through the expectation values
with respect to the physical states:
\begin{eqnarray*} 
  <\Psi|\hat{Q}_{kl}|\Psi>~=~<\Psi|\hat{\pi}^{kl}|\Psi>~=~0
\end{eqnarray*}
This is how the correspondence with the classical formulation emerges
in this framework.

%
Equation (\ref{annihilation}) completely specifies the dependence of
the wavefunctional on the variables $M_{kl}$, whereas the constraints
$G^{rot}_{i},~H_a$ and $\tilde{H}$ determine the $E_{i}^{a}$
dependence.  Note that in $\tilde{H}$, $M_{kl}$ appears only through
the corresponding constraint. Thus the full wavefunctional can be
written as:
\begin{eqnarray}\label{wavefunction}
  \Psi(M,E)~=~\tilde{\phi}(M-F)~\phi(E)
\end{eqnarray}
where $\tilde{\phi}(M-F)$ is a Gaussian functional of
$(M_{kl}-F_{kl})=Q_{kl}$. Thus the Gupta-Bleuler wavefunctional
differs from the one obtained through Dirac's procedure by the vacuum
of the oscillator in the Q space. The integral representation for the
inner product becomes:
\begin{eqnarray}\label{innerproduct}
  \int dM ~(\tilde{\phi} (M-F))^2~\int dE~\phi'^{*}(E) \phi(E)~=~\int dQ ~(\tilde{\phi} (Q))^2~\int dE~\phi'^{*}(E) \phi(E)
\end{eqnarray}
where we have used the fact that the Jacobian corresponding to the
change of variables from M to Q is identity. The Q integration can be
performed trivially, leaving only the E integral. This then becomes
equivalent to the reduced space integral as would be obtained by
Dirac's procedure, upto a normalisation.

Importantly, the above expression contains no delta-function
corresponding to the constraint in (\ref{annihilation}), which usually
appears as a projector in the inner product for first-class
constrained systems (see chapter-13 in \cite{teitelboim}, for
example). Also, note that the presence of the Gaussian functional
$\tilde{\phi}(M-F)$ in (\ref{innerproduct}) leads to normalisable
states in the $M_{kl}$ sector.  
\vspace{.7cm}
%

\subsection{Rescaling}

Next we proceed to perform the rescaling in the quantized phase space.
Y as in (\ref{ytime}) depends only on the operators corresponding to
the configuration variables ($E^{a}_{i}, M_{kl}$).  The new momenta
conjugate to $\hat{E^{a}_{i}}$ are thus given by :
\begin{eqnarray*}
  \hat{\omega}_{d}^{'0l}~\Psi' &~=~& e^{i\eta \int d^3x Y}~\hat{\omega}_{d}^{~0l}~e^{-i\eta\int d^3x Y}~\Psi'\\
  &~=~&\left(\frac{\delta}{\delta \hat{E}^{d}_{l}}~~-~\eta~\frac{\delta Y(\hat{E}^{a}_{i},\hat{M}_{jk})}{\delta \hat{E}^{d}_{l}}\right)~\Psi'\\
  &~=~&\left(\frac{\delta}{\delta \hat{E}^{d}_{l}}~-~\frac{\eta}{2}~\epsilon_{ijl}\hat{\omega}_{dij}~+~\eta~\left[~\frac{\hat{E}_{di}\hat{E}_{al}~-~\frac{1}{2}\hat{E}_{dl}\hat{E}_{ai}}{\sqrt{\hat{E}}}~\right]~\hat{t}^{a}_{i}\right)~\Psi'
\end{eqnarray*} 
%
where we have used the expression
\begin{eqnarray*}
  \frac{\delta Y}{\delta E^{d}_{l}}~&~=~&\epsilon_{ijk}\del_{b}\left(\frac{E_{ci}}{\sqrt{E}}\right)~\frac{\delta (\sqrt{E}E^{bj}E^{ck}~)}{\delta E^{d}_{l}}\\
  &~=~&~ \left(\frac{1}{2}\epsilon_{ijl}\omega_{dij}~-~\left[~\frac{E_{di}E_{al}~-~\frac{1}{2}E_{dl}E_{ai}}{\sqrt{E}}~\right]~t^{a}_{i}\right)
\end{eqnarray*}
with $t^{a}_{i}$ defined as :
\begin{eqnarray} \label{torsion}
  t^{a}_{i}~&=&~\epsilon^{abc}D_{b}V_{ci} \nonumber\\
  ~&=&~\epsilon^{abc}\del_{b}\left(\frac{E_{ci}}{\sqrt{E}}\right)~+~\frac{\sqrt{E}}{2}~[~\epsilon^{ijk}E^{a}_{k}\del_{b}E^{b}_{j}~+~E^{a}_{k}M_{ik}~-~E^{a}_{i}M_{kk}~]
\end{eqnarray} 

The new $\hat{\pi}_{kl}$'s are obtained as -
\begin{eqnarray*}
  \hat{\pi}'_{kl}~\Psi'&~=~&e^{i\eta \int d^3x Y}~\hat{\pi}_{kl}~e^{-i\eta \int d^3x Y}~\Psi'\\
  &~=~&\left(\frac{\delta}{\delta \hat{M}^{kl}}~-~\eta~\frac{\delta Y(\hat{E}^{a}_{i},\hat{M}_{jk})}{\delta \hat{M}^{kl}}\right)~\Psi'\\
  &~=~&\left(\frac{\delta}{\delta \hat{M}_{kl}}~+~\frac{\eta}{2}~\delta_{kl}\right)~\Psi'
\end{eqnarray*}

Note that this procedure goes through in presence of matter couplings
which lead to non-vanishing torsion.
 
As already mentioned, the expectation value of the constraint
(\ref{annihilation}) among the physical states in the $M_{kl}$ sector
(i.e., the states $\tilde{\phi}(M-F)$ in (\ref{wavefunction}) ) leads
to the relation-
\begin{eqnarray*}
  <\hat{M}_{kl}>_M~=~<F_{kl}(\hat{E_{i}^{a}})>_M~~~,
\end{eqnarray*}
which is the analogue of the classical constraint in (\ref{mkl}). To
emphasize, here torsion as an operator in (\ref{torsion}) does not
annihilate $\Psi$, rather its expectation value vanishes. This is to
be contrasted with the Dirac procedure where the torsion operator
vanishes `strongly'.
%

\newpage

{\bf New constraints} 
\vspace{.3cm}

The new constraints, which annihilate the rescaled wavefunctional
$\Psi'$, can be found by introducing the new momenta in the expressions
as given in (\ref{rot}) . We illustrate this for $\hat{G}^{rot}_{i}$
below.

The rotation constraint for the action in (\ref{ny}) containing the
Hilbert-Palatini and Nieh-Yan terms is given by :
\begin{eqnarray*}
  \hat{G}^{'rot}_{i}&~=&~e^{i\eta \int d^3x \hat{Y}}~\hat{G}^{rot}_{i}~e^{-i\eta \int d^3x \hat{Y}}\\
  &~=&~\eta~\del_{a}\hat{E}^{a}_{i}~+~\epsilon^{ijk}\hat{\omega}_{a}^{'0j}\hat{E}^{a}_{k}~-~\frac{\eta}{\sqrt{\hat{E}}}~\epsilon_{ijk}\hat{E}_{bj}\hat{t}^{b}_{k}
\end{eqnarray*}
Taking the expectation value with respect to the states
$\tilde{\phi}(M-F)$, we arrive at the familiar $SU(2)$ Gauss' law :
\begin{eqnarray*}
  <\hat{G}^{'rot}_{i}>_M~=~\eta \del_{a}\hat{E}^{a}_{i}~+~\epsilon^{ijk}\hat{A}_{a}^{j}\hat{E}^{a}_{k}
\end{eqnarray*}
where,
\begin{eqnarray*}
  \hat{A}_{d}^{l}~=~<\hat{\omega}_{d}^{'~0l}>_M~=~\hat{\omega}_{d}^{~0l}~-~\frac{\eta}{2}\epsilon^{ijl}\hat{\omega}_{dij}
\end{eqnarray*}
Without going into the detailed algebraic expressions of the remaining
constraints corresponding to (\ref{ny}) as they are not relevant for
our purpose here, we observe that they can be obtained in a similar
manner as shown for $\hat{G}^{rot}_{i}$.

\subsection{ Coherent state quantization}

We now demonstrate another approach, namely, the coherent state
quantization for constrained systems\cite{klauder}. Although this was
originally designed to develop an alternative path-integral
formulation using coherent states, here we use the essential idea to
enforce the appropriate `quantum' constraints. This would allow a
consistent rescaling formulation for gravity with or without matter.

Following the general construction developed by Klauder \cite{klauder}
, we seek the states for which
\begin{eqnarray}\label{firsteqn}
  <\Psi|\left((\hat{M}_{kl}-\hat{F}_{kl}(\hat{E}^{a}_{i}))^2+\hat{\pi}_{kl}^2\right)|\Psi>~=~<\Psi|(\hat{Q}_{kl}^2+\hat{\pi}_{kl}^2)|\Psi>~=~0
\end{eqnarray}
However, since we have $<\hat{A}^2>=(\Delta \hat{A})^2~+<\hat{A}>^2$
for any operator $\hat{A}$, equation (\ref{firsteqn}) cannot be
satisfied by any $\hat{A}$ with non-zero uncertainty $\Delta \hat{A}$.
Hence, as suggested in ref.\cite{klauder}, one has to modify the above
criterion as
\begin{eqnarray}\label{square} 
  <\Psi|(\hat{Q}_{kl}^2+\hat{\pi}_{kl}^2)|\Psi>~\leq~\lambda_{0}~\mathrm{(O(\hbar)})
\end{eqnarray}
where, $\lambda_0$ denotes the minimum eigenvalue in the spectrum of
the constraint operator. The modification, being of the order of
$\hbar$, is a purely quantum feature. The rest of the constraints
$G^{rot}_i,~H_a$ and $H$, as given by equation(\ref{rot}), are imposed
as they are on the physical states:
\begin{eqnarray}\label{neweqn}
\hat{G}^{rot}_{i}|\Psi>~=~\hat{H}_{a}|\Psi>~=~\hat{H}|\Psi>~=~0~~.
\end{eqnarray}

Now, there is a family of minimum uncertainty states, namely, the
canonical coherent states, defined as:
\begin{eqnarray}\label{coherent}
  |M,\pi>~=~e^{-iM\hat{\pi}}e^{i\pi\hat{M}}|\beta> \end{eqnarray}
where $M=<\hat{M}_{kl}>, ~\pi=<\hat{\pi}_{kl}>$ and $|\beta>$ is some
fiducial state for which $M=0, ~\pi=0$ (we supress the indices just to
simplify the notation). Among these, the one satisfying (\ref{square}) is the
coherent state for which $Q=(M-F(E))=0, ~\pi=0$, with
$F(E)=<\hat{F}_{kl}(\hat{E}^{a}_{i})>$.  Using (\ref{coherent}) , the explicit
form of this state reads:
\begin{eqnarray*} \Psi(M)~=~e^{-iF(E)\hat{\pi}}~\beta(M)~=~\beta(M-F(E))
\end{eqnarray*} where $\beta(M)$ is the fiducial state functional in a
representation diagonal in $\hat{M}_{kl}$.

 In this formulation, one can define a projection
operator P onto the physical Hilbert space, requiring the following
properties\cite{klauder}:
\begin{eqnarray*}
  P^{\dagger}=P,~~P^2=P
\end{eqnarray*}
In our case P (in the $M_{kl}$ sector) becomes simply
$|\Psi(M)><\Psi(M)|$ .

The full wavefunctional representing the physical subspace can thus be
written as:
\begin{eqnarray*}
  \Psi(M,E)~=~\beta(M-F)~ \phi(E)
\end{eqnarray*}
 The inner product in this
space reads:
\begin{eqnarray*}
  \int dM dE ~\Psi'^{*}(M,E)\Psi(M,E)~=~\int dQ ~\beta^{*}(Q)\beta(Q) \int dE~ \phi'^{*}(E)\phi(E) 
\end{eqnarray*}
As in the Gupta-Bleuler case, here also the $Q_{kl}$ (or, $M_{kl}$)
sector factors out leaving only the E integral in the product. In
particular, we can choose the fiducial state to be the oscillator
ground state in this case. Then this expression reproduces the
Gupta-Bleuler product as in (\ref{innerproduct}). Thus the two Hilbert
spaces are equivalent. The rescaling can now be implemented along the
lines of our previous discussion.

\section{Rescaling for any gauge choice}

Now we provide a brief outline of the rescaling procedure without
choosing any gauge. For non-zero $\chi_{i}$, the canonical coordinates
in the Lagrangian density in (\ref{hp}) are
$\tilde{\omega}_{a}^{~0i},~M_{kl}$ and $\chi_{i}$ where
$\tilde{\omega}_{a}^{~0i}$ is given by equation (\ref{tomega}) .

Y in (\ref{ybig}) can be rewritten as :
\begin{eqnarray} \label{Y}
  Y&~=&~\frac{1}{4}~\sqrt{E}\epsilon^{ijk}E^{a}_{j}E^{b}_{k}~\left[-~\chi_{i}~\del_{a}\left(\sqrt{E}\epsilon^{lmn}\epsilon_{bcd}~\chi_{l}E^{c}_{m}E^{d}_{n}\right)
    ~+~\del_{a}\left(\sqrt{E}\epsilon^{imn}\epsilon_{bcd}E^{c}_{m}E^{d}_{n}\right)\right]~\nonumber\\
  &-&\frac{1}{2}~M_{kk}~-~\frac{1}{2(1-\chi^2)}\chi_{k}\chi_{l}M_{kl}~-~\epsilon^{ijk}
  E_{j}^{a}\chi_{k}\del_{a}\chi_{i}~+~\chi_{k}G^{rot}_{k}
\end{eqnarray}
The structure of Y suggests that we can choose the representation to
be diagonal in the operators $\hat{E_{i}^{a}},~\hat{M_{kl}}$ and
$\hat{\chi_{i}}$ . In the above equation the last term involving
$\hat{G}^{rot}_i$ commutes with all other remaining terms and acts
trivially on $|\Psi>$ to give zero. Hence this term can be ignored at
this stage itself.

One can follow exactly the same procedure as earlier to define a
suitable physical subspace using either the Gupta-Bleuler or the
coherent state method, and then find the new set of canonical
operators through the rescaling. Thus, the new momenta
$~\tilde\omega_{a}^{'0i}$ conjugate to $E^{a}_{i}$ read:
\begin{eqnarray} \label{Omega}
  \tilde{\omega}_{d}^{'0l}&~=&~\omega_{d}^{(\eta)~0l}~-~\chi_{j}\omega_{d}^{(\eta)~lj}~-~\eta\epsilon^{ikl}\chi_{k}\del_{d}\chi_{i} \nonumber \\
  &&~+~\eta~\left(~\frac{E_{ai}E_{dl}}{2
      \sqrt{E}}~+~\sqrt{E}\epsilon^{ilk}\epsilon_{abd}E^{b}_{k}\right)(t^{a}_{i}~-~\chi_{i}t^{a}_{0})
\end{eqnarray}
where we have defined :
\begin{eqnarray*}
  \omega_{d}^{(\eta)~0l}&~=~&\omega_{d}^{~0l}~-~\frac{\eta}{2}~\epsilon^{jkl} \omega_{djk}~~,~~\omega_{d}^{(\eta)~lj}~=~\omega_{d}^{~lj}~-~\eta~\epsilon^{jkl}\omega_{d0k}~~,\\
  t^{a}_{i}&~=~&\epsilon^{abc}~D_{b}V_{ci}~=~\epsilon^{abc}~[~\del_{b}V_{ci}~+~\omega_{bij}V_{c}^{j}~-~\omega_{b}^{0i}V_{c0}~]~~,\\
  t^{a}_{0}&~=~&\epsilon^{abc}~D_{b}V_{c0}~=~\epsilon^{abc}~[~\del_{b}V_{c0}~-~\omega_{b}^{0i}V_{c}^{i}~]
  \end{eqnarray*}
  As is evident, we recover the new momenta in time gauge when
  $\chi_{i}=0$ and $\zeta_{i}=\del_{a}E^{a}_{i}$ .

  The momenta corresponding to $M_{kl}$ and $\chi_{i}$ transform as
  follows:
\begin{eqnarray*}
  \pi'_{kl}~&=&~\pi_{kl}~+~\frac{\eta}{2}\left(\delta_{kl}+\frac{\chi_k\chi_l}{1-\chi^2}\right)\nonumber\\
  {\zeta}'^{i}~&=&~\zeta^{i}~-~\eta\epsilon^{ijk}E_{j}^{a}\left[\del_{a}\chi_{k}~-~\sqrt{E}E_{k}^{b}\chi_{m}\del_{a}\left(\frac{E_{b}^{m}}{\sqrt{E}}\right)\right] ~+~\frac{\eta}{1-\chi^2}\left(\delta_{ij}+\frac{\chi_i\chi_j}{1-\chi^2} \right)\chi_k M^{jk}
\end{eqnarray*}
The new set of constraints, which annihilate the rescaled
wavefunctional $\Psi'$ can now be obtained in a manner as demonstrated
for the time-gauge fixed theory.



%
%

\section{Conclusion} 

We have illustrated how to arrive at the canonical formulation
corresponding to the action containing the Hilbert-Palatini and
Nieh-Yan terms starting from the Hilbert-Palatini canonical theory
through a generic rescaling procedure. The constraint operators,
through their action on the physical states, reproduce the real
Ashtekar-Barbero formulation. 

As it turns out, one cannot invoke such a rescaling to obtain the
Ashtekar-Barbero constraints if the Hilbert-Palatini second-class
constraints are eliminated before quantization, as in Dirac's method.
Here we have provided a remedy to this problem by using alternative
approaches, namely, the Gupta-Bleuler and the coherent state
quantizations. These two cases result in the same physical Hilbert
space. Also, here the torsional degrees of freedom as associated with
the second-class constraints emerge as relevant canonical operators
through $\hat{M}_{kl}$ and $\hat{\pi}^{kl}$. To emphasize, these do
not appear in the Dirac-quantized phase space where the second-class
constraints are eliminated beforehand.

As both the quantization methods lead to a non-vanishing rescaling
functional, they apply to any arbitrary matter coupling.  When such
couplings lead to nonzero torsion (e.g. fermion coupled to gravity),
one can obtain the new canonical constraints by writing the rescaling
functional Y in terms of the geometric variables (i.e., tetrads and
spin connections).  Using the connection equation of motion to write Y
in terms of matter fields there becomes purely optional.  Thus our
analysis provides a complete topological interpretation of the
Barbero-Immirzi parameter in a quantum framework, whether or not
matter is coupled to gravity. 

We have also demonstrated that the rescaling can be carried out
without choosing any particular gauge (e.g., time gauge). Thus the
appearance of $\eta$ as a topological parameter in this quantum
description is not an artifact of some special gauge choice, as also
shown in ref.\cite{kaul} from a classical perspective.

The construction here clearly shows that $\eta$ can manifest itself as
a vacuum angle through a QCD like rescaling provided the
representation is chosen to be diagonal in the densitized triad
operators. Also, both the quantization approaches result in a
well-defined operator corresponding to the `large gauge
transformations'. Thus, our analysis might be particularly relevant in
the context of a path-integral quantization, which is the most natural
arena to investigate the effects coming from a potential
non-perturbative vacuum structure underlying such transformations.
\vspace{.8cm}
%


{\bf Acknowledgements:} 

Comments of Prof. G. Date, Prof. R. Kaul and Alok Laddha are
gratefully acknowledged. It is also a pleasure to thank Martin
Bojowald (who pointed out ref.\cite{klauder}) and Thomas Thiemann for
general discussions during the `Loops '09' meeting, Beijing.

\end{document}